\documentclass[pra,a4paper,notitlepage]{revtex4-1}
\pdfoutput=1

\usepackage[verbose=true]{microtype}
\usepackage[utf8]{inputenc}
\usepackage[intlimits]{amsmath}
\usepackage{amssymb,amsfonts,amsthm}
\usepackage{enumerate}
\usepackage[capitalise]{cleveref}
\usepackage{graphicx}
\usepackage{caption}
\usepackage{subcaption}

\theoremstyle{plain}

\DeclareMathOperator{\sech}{sech}
\DeclareMathOperator{\sgn}{sgn}
\renewcommand{\epsilon}{\varepsilon}

\newcommand{\be}{\begin{eqnarray}}
\newcommand{\ee}{\end{eqnarray}}
\newcommand{\ba}{\begin{array}}
\newcommand{\ea}{\end{array}}

\newcommand{\ben}{\begin{enumerate}}
\newcommand{\een}{\end{enumerate}}
\newcommand{\bi}{\begin{itemize}}
\newcommand{\ei}{\end{itemize}}

\begin{document}

\title{Generalization of Legendre functions applied to Rosen-Morse scattering states}

\author{F.~L.~Freitas}
\email{felipelopesfreitas@gmail.com}

\begin{abstract}
A generalization of associated Legendre functions is proposed and used to describe the scattering states of the Rosen-Morse potential. The functions are then given explicit formulas in terms of the hypergeometric function, their asymptotic behavior is examined and shown to match the requirements for states in the regions of total and partial reflection. Elementary expressions are given for reflection and transmission coefficients, and an integral identity for the generalized Legendre functions is proven, allowing the calculation of the spectral measure of the induced integral transform for the scattering states. These methods provide a complete classical solution to the potential, without need of path integral techniques. \end{abstract} 

\maketitle

\section{Introduction}
\label{sec:introduction}

It is well known\cite{Natanzon1979} that the hypergeometric function provides closed form solutions to the Schr\"odinger equation for a large number of potentials. Nevertheless, a full classification of these potentials and their solutions remains difficult, in part because the orthogonality of their states cannot be described in terms of classical Sturm-Liouville theory, since their differential equations possess singularities.

Perhaps the most famous instance of this theoretical phenomenon is in the radial differential equation for the hydrogen atom. See, for example, exercise 13.2.11 of Arfken and Weber\cite{Arfken2005}, where although Laguerre polynomials appear in the potential wave functions, the orthogonality relations they obey are unrelated to the one actually derived in the construction of said polynomials.

For this reason, generaliations of the classical theory are often proposed, leading to new insights about the mathematics behind these analytically solvable systems. One example is the rediscovery of Romanovski polynomials and their finite orthogonality\cite{Weber2007} in the bound states of a hypergeometric potential\cite{Compean2006}.

Of all these solvable potentials, one of the most useful and insteresting is the one proposed by Rosen and Morse\cite{Rosen1932}, given by the expression 
\begin{equation}
V(x) = -\frac{\hbar^2\alpha(\alpha+1)}{2m\Delta^2}\sech^2\left(\frac{x}{\Delta}\right) + \frac{\hbar^2\beta}{m\Delta^2}\tanh\left(\frac{x}{\Delta}\right).
\label{eq:rosenmorsefull}
\end{equation}

Although initially proposed to study the properties of molecules, the potential soon drew attention to its mathematical properties. Not only it describes a well trapping a finite number of bound states, but it's also like a smooth step potential, possessing a zone where particles are totally reflected, and another where they have a chance of being transmitted with changes in frequency.

The Rosen-Morse potential can be studied in the context of path integrals\cite{Kleinert1992,Lin1997,Grosche1989}, where its propagator is derived by relating the problem to motion in a manifold and applying transformations. It is also a generalization of the modified P\"oschl-Teller potential\cite{Flugge2012}, the Woods-Saxon potential\cite{Woods1954} and the Morse-Feshbach potential\cite{Morse1953}.

It has been noticed recently\cite{Freitas2018} that the operator linking adjacent states in the Rosen-Morse potential is nonlocal, and needs tools from fractional calculus for its construction. This is in stark contrast to its symmetric form, where a differential operator does exist and is obtainable from properties of associated Legendre functions\cite{Dong2002}, suggesting that the orthogonality of the Rosen-Morse eigenstates is in a sense deeper than the modified P\"oschl-Teller and that of the Coulomb potential.

In this work, we propose that the most natural description of these states is not in terms of the plain hypergeometric function, but of closely related special functions. These functions are the solution of a differential equation very similar to that of Legendre functions, and elegantly describe both the scattering and bound states of the Rosen-Morse potential.

Since the normalization constants of the bound states have already been derived previously\cite{Freitas2018}, we focus mostly on scattering states, calculating reflection and transmission coefficients by analyzing the asymptotic behavior of the functions, and deriving the normalization of the continuum spectrum with a generalization of an integral identity for associated Legendre functions reported by Magnus, Oberhettinger and Soni\cite{Magnus1966} and applied to quantum mechanical problems by Grosche\cite{Grosche1990}. This normalization is then used to evaluate the spectral measure of the integral transform induced by these functions, and we arrive at much simpler expressions than those obtained with path integral methods\cite{Grosche1998}.

This paper is organized as follows.  In Section \ref{sec:genleg} the generalized Legendre functions are defined, shown to be solutions to the Schr\"odinger equation for the Rosen-Morse potential, and expressed in terms of the hypergeometric function. In Section \ref{sec:asymp} their asymptotic behavior is elucidated, and it's shown that they can oscillate at different frequencies or even decay or grow exponentially at different rates at $x\to+\infty$ and at $x \to-\infty$. In section \ref{sec:states}, we exploit our asymptotic analysis to map the parameters of the generalized functions to the energy of Rosen-Morse scattering states, while discussing briefly how one could also use them to find bound states. In section \ref{sec:coefs}, we calculate reflection and transmission coefficients for states with energy above the potential barrier, using several identities to simplify the final expression. In section \ref{sec:integral}, we prove a useful integral identity for the generalized Legendre functions. In section \ref{sec:norm}, the identity is used to evaluate the continuum normalization of the scattering states and the spectral measure for their integral transform. Finally, in section \ref{sec:concl}, the conclusions are outlined.

\section{Generalized Legendre functions}
\label{sec:genleg}

	Before we start, let us briefly state the conventions for usual associated Legendre functions, taken from Gradshteyn and Ryzhik\cite{Gradshteyn2014}. An associated Legendre function is a particular solution to the differential equation
	
	\begin{equation}\label{eq:assocdiff}
	(1-z^2)\frac{d^2u}{dz^2} - 2z\frac{du}{dz} + \left[\nu(\nu+1) - \frac{\mu^2}{1-z^2}\right]u = 0,
	\end{equation}
where $\mu$ and $\nu$ are understood as arbitrary complex parameters.

	These functions are defined in the complex plane with a branch cut in the interval $(-\infty,1)$. In our case, though, we are interested in solutions in the interval $(-1,1)$, and these are constructed by averaging the previously mentioned complex function above and below the cut, which also solve \eqref{eq:assocdiff}. These functions \emph{on the cut} are the ones we are interested in generalizing.

	With this in place, let us consider functions $D^{\mu,\eta}_\nu(x)$, which satisfy a more general equation than \eqref{eq:assocdiff}, given by
	
	\begin{equation}\label{eq:defeq}
	(1-x^2)\frac{d^2D^{\mu,\eta}_\nu}{dx^2}-2x\frac{dD^{\mu,\eta}_\nu  }{dx} + \left[\nu(\nu+1) - \frac{\mu^2+2\mu\eta x + \eta^2}{1-x^2}\right]D^{\mu,\eta}_\nu   = 0.
	\end{equation}
	
	The motivation for considering this equation becomes clearer when we perform the change of variables $x = \tanh(v)$. Writing $\psi=D^{\mu,\eta}_\nu  $, we can do the transformation
	
	\begin{equation}
	\frac{d\psi}{dx} = \cosh^2 v\frac{d\psi}{dv}
	\end{equation}
	
	\begin{equation}
	\frac{d^2\psi}{dx^2} = \cosh^2 v \frac{d}{dv} \left(\cosh^2 v\frac{d\psi}{dv}\right) = \cosh^4 v \frac{d^2\psi}{dv^2} + 2\sinh v \cosh^3v \frac{d\psi}{dv}
	\end{equation}
	
	Substituting on \eqref{eq:defeq}, we obtain
	
	\begin{equation}
	\cosh^2 v \frac{d^2\psi}{dv^2} + \left[\nu(\nu+1) - 2\mu\eta\sinh v\cosh v - (\mu^2+\eta^2)\cosh^2 v\right]\psi = 0.
	\end{equation}
	
	Another way of writing this differential equation is
	
	\begin{equation}\label{eq:rosenmorse}
	\psi'' + [\nu(\nu+1)\sech^2 v - 2\mu\eta \tanh v - \mu^2 - \eta^2]\psi = 0,
	\end{equation}
and comparing it with the adimensional form of \eqref{eq:rosenmorsefull}, obtained with $\hbar=2m=\Delta=1$:

	\begin{equation}\label{eq:rosenmorsenodim}
	\psi'' + [\alpha(\alpha+1)\sech^2 x - 2\beta \tanh x +E]\psi = 0,
	\end{equation}		
it is readily seen that they are equivalent by means of the substitutions $E=-\mu^2-\eta^2$, $\mu\eta = \beta$ and $\nu=\alpha$.

	Equation \eqref{eq:rosenmorse} can be transformed into the hypergeometric differential equation. We will do this by closely following the method used in a previous work\cite{Freitas2018}, but there are important differences, since we want to reduce to associated Legendre functions when $\eta=0$. The basic idea is to employ the \emph{ansatz}
	
	\begin{equation}
	\psi(v) = e^{-av}\cosh^{-b} v F(v).
	\end{equation}
	
	The equation changes to\cite{Freitas2018}:
	
	\begin{equation}\label{eq:hyperv}
	F'' - 2(a+b\tanh v)F' + \left\lbrace \left[\nu(\nu+1)-b(b+1)\right]\sech^2 v +(2ab-2\mu\eta)\tanh v + a^2 + b^2 - \mu^2 -\eta^2 \right\rbrace F  = 0.
	\end{equation}
	
	The behavior at $v\to+\infty$ is

	\begin{equation}\label{eq:asymplus}
	F'' - 2(a+b)F' + \left[(a+b)^2-2\mu\eta - \mu^2 -\eta^2\right]F = 0,
	\end{equation}
and at $v\to-\infty$:

	\begin{equation}\label{eq:asymminus}
	F'' - 2(a-b)F' + \left[(a-b)^2+2\mu\eta - \mu^2 -\eta^2\right]F = 0.
	\end{equation}
	
	Parameters $\mu$ and $\eta$ are chosen so that the factors multiplying $F$ at \eqref{eq:asymplus} and \eqref{eq:asymminus} vanish. In order to obtain the correct Legendre functions at the $\eta=0$ limit, we must use
	
	\begin{equation}
	a = -\mu, \qquad b = -\eta.
	\end{equation}
	
	Using these values for $a$ and $b$ we get many cancellations on \eqref{eq:hyperv}, and changing to $u=(1+\tanh v)/2$, we find\cite{Freitas2018}:
	
	\begin{equation}
	u(1-u)\frac{d^2F}{du^2} + [\mu-\eta+1-2(1-\eta)u]\frac{dF}{du} + [\nu(\nu+1)+\eta(1-\eta)]F = 0.
	\end{equation}
		
	This is the hypergeometric equation, typically written as\cite{Abramowitz1965}

	\begin{equation}\label{eq:hyperdiff}
	u(1-u)F'' + [t-(r+s+1)u]F' - rsF = 0.
	\end{equation}
	
	For our problem, the parameters are
	
	\begin{equation}
	r = -\eta-\nu, \qquad s = -\eta + \nu + 1, \qquad t = -\eta+\mu+1.
	\end{equation}
	
	Equation \eqref{eq:hyperdiff} is a second order differential equation, and we're interested in solutions in the interval $[0,1]$. Its set of solutions forms a two-dimensional vector space, and there are several ways to express the general form. Our strategy will be to pick a solution of \eqref{eq:hyperdiff} and define our generalized Legendre function $D^{\mu,\eta}_\nu  (x)$ in terms of it, so that when $\eta=0$ we reduce to the usual associated Legendre function. It turns out that, to achieve this goal, the solution of \eqref{eq:hyperdiff} we need to choose is\cite{Abramowitz1965}:
	
	\begin{equation}
	F(r,s;r+s+1-t;1-u) = F\left(-\nu-\eta,\nu+1-\eta;1-\mu-\eta; (1-x)/2\right)
	\end{equation}
	
	In view of the hyperbolic relations
	
	\begin{equation}
	x = \tanh v = \frac{e^{2v} - 1}{e^{2v} + 1} \Rightarrow e^{2v}(x-1) = -1-x \Rightarrow e^{2v} = \frac{1+x}{1-x},
	\end{equation}
and

	\begin{equation}
	\cosh v = \sqrt{\frac{1}{1-\tanh^2 v}} = \sqrt{\frac{1}{1-x^2}},
	\end{equation}
we can finally write the definition for our generalized Legendre function:

	\begin{equation}\label{eq:defgen}
	D^{\mu,\eta}_\nu  (x) = (1-x^2)^{-\frac{\eta}{2}}\left(\frac{1+x}{1-x}\right)^{\frac{\mu}{2}}F\left(-\nu-\eta,\nu+1-\eta;1-\mu-\eta; (1-x)/2\right).
	\end{equation}
	
	If we compare this with the definition of usual associated Legendre functions\cite{Gradshteyn2014}:
	
	\begin{equation}\label{eq:defleg}
	P^\mu_\nu(x) = \frac{1}{\Gamma(1-\mu)}\left(\frac{1+x}{1-x}\right)^{\frac{\mu}{2}}F(-\nu,\nu+1;1-\mu;(1-x)/2)
	\end{equation}
	
	Notice that, when $\beta = 0$, \eqref{eq:defgen} reduces to
	
	\begin{equation}
	D^{\mu,0}_\nu(x) = \Gamma(1-\mu)P^\mu_\nu(x).
	\end{equation}

\section{Asymptotic analysis}
\label{sec:asymp}

Since our wavefunctions will have the form $D^{\mu,\eta}_\nu(\tanh x)$, let's analyze the asymptotics of this function in $x\to+\infty$.
	
	When we send $x\to\tanh x$ in \eqref{eq:defgen}, we get
	
	\begin{equation}
	\frac{1+\tanh x}{1-\tanh x} = e^{2x}, \qquad 1-\tanh^2 x = \sech^2 x = \cosh^{-2} x. 
	\end{equation}
	
	Since $\cosh x \to e^x/2$ at $x\to+\infty$ and the hypergeometric factor goes to $1$ for an argument of $0$, we obtain
	
	\begin{equation}\label{eq:limplus}
	\lim_{x\to+\infty}D^{\mu,\eta}_\nu(\tanh x) = 2^{-\eta}e^{(\mu+\eta)x}.
	\end{equation}
	
	Because of this, $\mu+\eta$ cannot have a positive real part, or else we don't get a normalizable solution.
	
	To find the asymptotics at $x\to-\infty$, use the hypergeometric linear transformation formula\cite{Abramowitz1965}
	
	\begin{equation}\label{eq:lintransf}
	\begin{aligned}
	F(a,b;c;z) = &\frac{\Gamma(c)\Gamma(c-a-b)}{\Gamma(c-a)\Gamma(c-b)}F(a,b;a+b-c+1;1-z) \\
	&+(1-z)^{c-a-b}\frac{\Gamma(c)\Gamma(a+b-c)}{\Gamma(a)\Gamma(b)}F(c-a,c-b;c-a-b+1;1-z).
	\end{aligned}
	\end{equation}
	
	For our generalized Legendre function:
	
	\begin{equation}\label{eq:shifted}
	\begin{aligned}
	D^{\mu,\eta}_\nu(x) = &(1-x^2)^{-\frac{\eta}{2}}\left(\frac{1+x}{1-x}\right)^{\frac{\mu}{2}}\bigg[\frac{\Gamma(1-\mu-\eta)\Gamma(-\mu+\eta)}{\Gamma(1-\mu+\nu)\Gamma(-\mu-\nu)}F(-\nu-\eta,\nu+1-\eta;1+\mu-\eta;(1+x)/2) \\
	&+\left(\frac{1+x}{2}\right)^{-\mu+\eta}\frac{\Gamma(1-\mu-\eta)\Gamma(\mu-\eta)}{\Gamma(-\nu-\eta)\Gamma(\nu+1-\eta)}F(1-\mu-\nu,\nu-\mu;1-\mu+\eta;(1+x)/2)\bigg].
	\end{aligned}
	\end{equation}
	
	Notice that, when we send $x\to\tanh x$ in \eqref{eq:shifted}, in the limit $x\to-\infty$, the hypergeometric factors go to $1$, and the factor $(1+\tanh x)/2 \to e^{2x}$, because we have the limit
	
	\begin{equation}
	\lim_{x\to-\infty} \frac{1+\tanh x}{e^{2x}} = 2.
	\end{equation}
	
	With $\cosh x \to e^{-x}/2$ at $x\to-\infty$, we have
	
	\begin{equation}\label{eq:limminus}
	\lim_{x\to-\infty} D^{\mu,\eta}_\nu(x) = 2^{-\eta}\left[ \frac{\Gamma(1-\mu-\eta)\Gamma(-\mu+\eta)}{\Gamma(1-\mu+\nu)\Gamma(-\mu-\nu)}e^{(\mu-\eta)x} + \frac{\Gamma(1-\mu-\eta)\Gamma(\mu-\eta)}{\Gamma(-\nu-\eta)\Gamma(\nu+1-\eta)} e^{-(\mu-\eta)x}\right].
	\end{equation}

\section{Potential eigenstates}
\label{sec:states}

The goal of this section is to make formulas \eqref{eq:limplus} and \eqref{eq:limminus} more physical by relating the parameters in the generalized Legendre function with those of the Rosen-Morse potential. As stated before, we must perform the following substitution on \eqref{eq:rosenmorsenodim}:

	\begin{equation}
	\mu^2+\eta^2 = -E, \qquad \mu\eta = \beta.
	\end{equation}
	
	If we are looking for bound states, then we must have $E < -2\beta$. In this case, we can write $E=-2\beta - b^2$. The parameters $\mu$ and $\eta$ are both real and are given by
	
	\begin{equation}
	\mu = \frac{b - \sqrt{4\beta + b^2}}{2}, \qquad \eta = \frac{-b - \sqrt{4\beta + b^2}}{2}.
	\end{equation}
	
	Notice that we must choose $\mu$ and $\eta$ to be both negative, or else the asymptotic behavior at $+\infty$ from \eqref{eq:limplus} would not lead to a normalizable state. Besides that, the limit at $-\infty$ from \eqref{eq:limminus} has two contributions from $e^{(\mu-\eta)x}$ and $e^{-(\mu-\eta)x}$. Since in our case $\mu-\eta>0$, the $e^{-(\mu-\eta)x}$ term needs to vanish, and we can only do this by placing the arguments of the gamma functions at the poles at nonpositive integers.
	
	From this, the discrete spectrum can be obtained and the wavefunctions, together with their normalization constants, obtained from properties of Jacobi polynomials. Details can be found in previous work\cite{Freitas2018}. Currently, we are interested in the continuum spectrum, so we discuss scattering states next.
	
	There are two scattering regimes, the first with $-2\beta < E < 2\beta$, where we scatter below the barrier and get total reflection, and the regime with $E>2\beta$, where we scatter above the barrier and can have nonzero transmission.
	
	For the first regime, let's set $E=-2\beta+k^2$. The parameters assume the values
	
	\begin{equation}\label{eq:scattering}
	\mu = -\frac{\sqrt{4\beta-k^2}}{2} + \frac{ik}{2}, \qquad \eta = -\frac{\sqrt{4\beta-k^2}}{2} - \frac{ik}{2}.
	\end{equation}
	
\begin{figure}[htp]
\hspace*{\fill}%
\begin{minipage}[t]{.45\textwidth}
\centering
\vspace{0pt}
  \includegraphics[width=\linewidth]{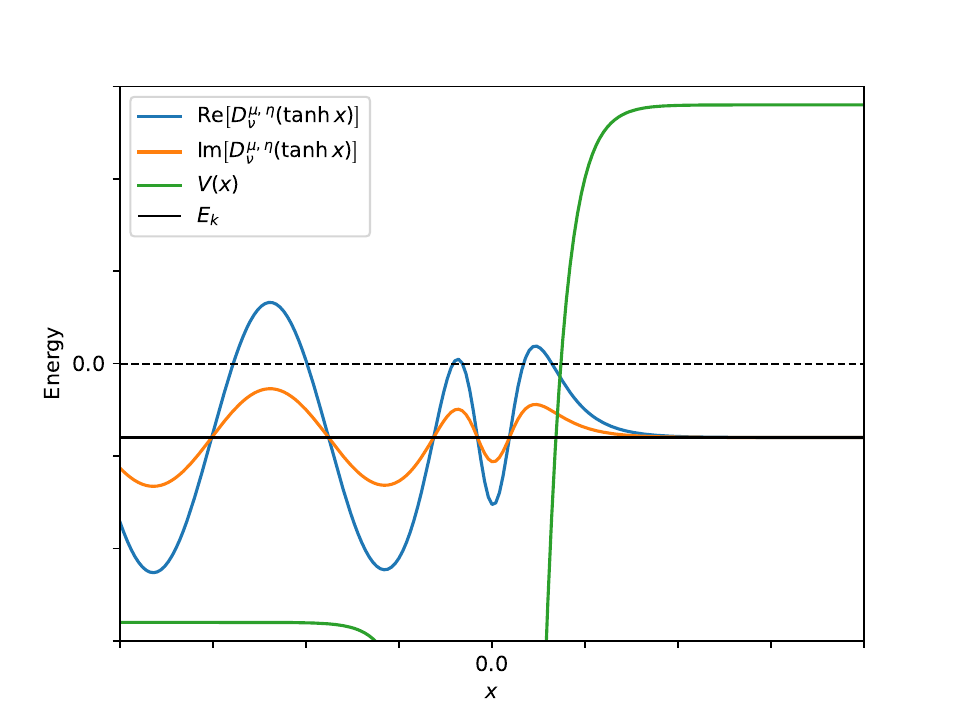}
  \captionof{figure}{Rosen-Morse scattering state $D^{\mu,\eta}_\nu(\tanh x)$ below the energy barrier. The solution is a distorted wave close to the well, decays into the classically forbidden region as $x\to+\infty$ and shows a wave pattern typical of a wave interfering with its reflection as $x\to-\infty$.}
  \label{fig:below}
\end{minipage}%
\hfill
\begin{minipage}[t]{.45\textwidth}
\centering
\vspace{0pt}
  \includegraphics[width=\linewidth]{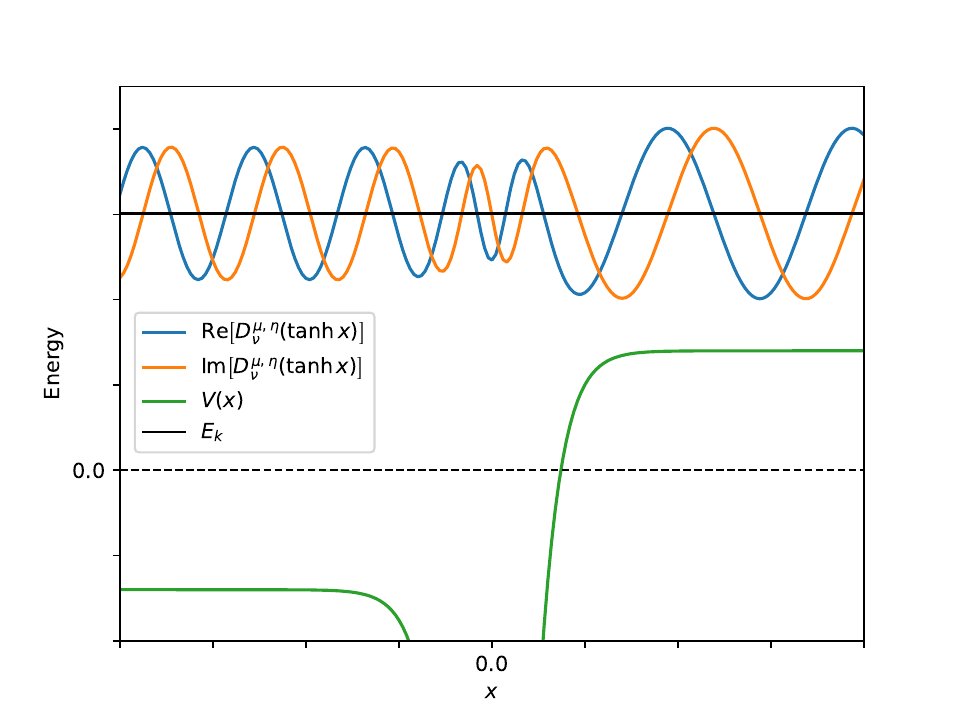}
  \captionof{figure}{Rosen-Morse scattering state $D^{\mu,\eta}_\nu(\tanh x)$ above the energy barrier. The solution is a distorted wave close to the well. The frequency at $x>0$ is lower than that at $x<0$. A large portion of the wave is transmitted, so the amplitude increases at $x>0$ to conserve probability current at lower frequency.}
  \label{fig:above}
\end{minipage}
\hspace*{\fill}%
\end{figure}

	We see that both $\mu$ and $\eta$ have a real and an imaginary part, and are complex conjugates of each other. There are two degenerate states for this energy, one for $k$ and the other for $-k$, and they will contain the information regarding the phase shift caused by the reflection depending on the polarization of the incoming wave. Since $\mu+\eta<0$, all states decay to zero at $x\to+\infty$, as expected. The qualitative behavior of solutions in this regime is plotted in Figure~\ref{fig:below}.
	
	In the second regime, above the barrier but also with $E=-2\beta+k^2$, the square root in \eqref{eq:scattering} gives us an imaginary number, and we have
	
	\begin{equation}\label{eq:scatpos}
	\mu = \frac{i\sgn(k)\sqrt{k^2-4\beta}}{2} + \frac{ik}{2}, \qquad \eta = \frac{i\sgn(k)\sqrt{k^2-4\beta}}{2} - \frac{ik}{2}.
	\end{equation}
	
	Now, both $\mu$ and $\eta$ are purely imaginary. At $x\to+\infty$, our solution oscillates with a single mode $e^{i\sgn(k)\sqrt{k^2-4\beta}}$. At $x\to-\infty$, there are two modes, $e^{ikx}$ and $e^{-ikx}$. The qualitative behavior is plotted in Figure~\ref{fig:above}. Notice that our choice of the sign of the square root is dictated by the sign of $k$, as emphasized by the $\sgn$ function. This is important to guarantee that the states corresponding to $k$ and $-k$ are orthogonal.

\section{Reflection and transmission coefficients}
\label{sec:coefs}
	
	With the asymptotic analysis done, we are ready to evaluate reflection and transmission coefficients. We need a setup like
	
	\begin{equation}\label{eq:upbarrier}
	\psi_k(x) = \begin{cases}
		Ae^{ikx} + Be^{-ikx} & x \to -\infty \\
		Ce^{i\sqrt{k^2-4\beta}x} & x \to +\infty
	\end{cases}.
	\end{equation}
	
	Notice that this is exactly the asymptotics of our eigenstate, and it corresponds to particles going from left to right, and being either reflected or transmitted through the potential. The coefficients are given by

	\begin{equation}
	R = \frac{|B|^2}{|A|^2}, \qquad T = \frac{\sqrt{k^2-4\beta}}{k}\frac{|C|^2}{|A|^2} = \frac{(\mu+\eta)}{(\mu-\eta)}\frac{|C|^2}{|A|^2}.
	\end{equation}		
	
	Although it is not intuitive, if we instead set up a wave coming from the right, the reflection and transmission coefficients would be the same, even though the particle would see a drop in the potential. This follows from very general properties of the Schrödinger equation\cite{Landau2013}.
	
	Let's first consider the regime below the barrier, where transmission is trivially $T_{below}=0$. We apply \eqref{eq:limminus}, with $\mu-\eta=ik$ from \eqref{eq:scattering}. To evaluate the reflection coefficient we first divide the amplitude multiplying $e^{-ikx}$ by the amplitude of $e^{ikx}$.
	
	\begin{equation}\label{eq:refbelow}
	\frac{B}{A} = \frac{\Gamma(\mu-\eta)\Gamma(1-\mu+\nu)\Gamma(-\mu-\nu)}{\Gamma(-\nu-\eta)\Gamma(\nu+1-\eta)\Gamma(-\mu+\eta)}
	\end{equation}
	
	In view of \eqref{eq:scattering}, we can write $\mu=a+bi$ and $\eta=a-bi$ for real $a$ and $b$. We also substitute $\nu=\alpha$, which is also real.
	
	\begin{equation}
	\frac{B}{A} = \frac{\Gamma(2bi)\Gamma(1-a-bi+\alpha)\Gamma(-\alpha-a-bi)}{\Gamma(-\alpha-a+bi)\Gamma(\alpha+1-a+bi)\Gamma(-2bi)}
	\end{equation}
	
	Notice that $B/A = 1/(B/A)^*$, and since we have $R_{below}=\frac{|B|^2}{|A|^2}$, we have
	
	\begin{equation}
	R_{below} = 1, \qquad T_{below} = 0,
	\end{equation}
as expected.

	Now, for the more interesting process, where we scatter above the barrier. In this situation, the reflection coefficient is a formula identical to \eqref{eq:refbelow}, but instead we substitute \eqref{eq:scatpos}. When we calculate the magnitude squared, we must remember that both $\mu$ and $\eta$ are purely imaginary.
	
	\begin{equation}
	R = \frac{\Gamma(\mu-\eta)\Gamma(1-\mu+\alpha)\Gamma(-\mu-\alpha)}{\Gamma(-\alpha-\eta)\Gamma(\alpha+1-\eta)\Gamma(-\mu+\eta)}\frac{\Gamma(-\mu+\eta)\Gamma(1+\mu+\alpha)\Gamma(\mu-\alpha)}{\Gamma(-\alpha+\eta)\Gamma(\alpha+1+\eta)\Gamma(\mu-\eta)}.
	\end{equation}
	
	Cancelling several terms and applying the gamma reflection formula
	
	\begin{equation}\label{eq:gammaid}
	\Gamma(z)\Gamma(1-z) = \frac{\pi}{\sin(\pi z)},
	\end{equation}
we write

	\begin{equation}
	R = \frac{\sin(\pi(-\alpha+\eta))\sin(\pi(-\alpha-\eta))}{\sin(\pi(-\alpha+\mu))\sin(\pi(-\alpha-\mu))}.
	\end{equation}
	
	Since $\alpha$ is real, and $\mu$ and $\eta$ are imaginary, we can apply the trigonometric identity
	
	\begin{equation}\label{eq:trigid}
	\sin(x+iy)\sin(x-iy) = \sin^2(x)+\sinh^2(y).
	\end{equation}
	
	Therefore,
	
	\begin{equation}
	R = \frac{\sin^2(\pi\alpha)+\sinh^2\left(\frac{\pi}{2}(k-\sqrt{k^2-4\beta})\right)}{\sin^2(\pi\alpha)+\sinh^2\left(\frac{\pi}{2}(k+\sqrt{k^2-4\beta})\right)}.
	\end{equation}
	
	To evaluate transmission we divide the coefficient multiplying the $e^{i\sqrt{k^2-4\beta}x}$ in \eqref{eq:limplus} by the coefficient of $e^{ikx}$ in \eqref{eq:limminus}:
	
	\begin{equation}
	\frac{C}{A} = \frac{\Gamma(1-\mu+\alpha)\Gamma(-\mu-\alpha)}{\Gamma(1-\mu-\eta)\Gamma(-\mu+\eta)}.
	\end{equation}
	
	The transmission coefficient is then
	
	\begin{equation}
	T = \frac{\sqrt{k^2-4\beta}}{k}\frac{|C|^2}{|A^2|} = \frac{(\mu+\eta)}{(\mu-\eta)}\frac{\Gamma(1-\mu+\alpha)\Gamma(-\mu-\alpha)}{\Gamma(1-\mu-\eta)\Gamma(-\mu+\eta)}\frac{\Gamma(1+\mu+\alpha)\Gamma(\mu-\alpha)}{\Gamma(1+\mu+\eta)\Gamma(\mu-\eta)}.
	\end{equation}
	
	To simplify it we need \eqref{eq:gammaid} and the gamma shift identity
	
	\begin{equation}
	\Gamma(z+1) = z\Gamma(z),
	\end{equation}
we have

	\begin{equation}
	T = \frac{\sin(\pi(\mu-\eta))\sin(\pi(-\mu-\eta))}{\sin(\pi(\mu-\alpha))\sin(\pi(-\mu-\alpha))}.
	\end{equation}
	
	Using \eqref{eq:trigid} again:
	
	\begin{equation}
	T = \frac{\sinh(\pi k)\sinh\left(\pi\sqrt{k^2-4\beta}\right)}{\sin^2(\pi\alpha)+\sinh^2\left(\frac{\pi}{2}(k+\sqrt{k^2-4\beta})\right)}.
	\end{equation}
	
	As expected, we have $R+T=1$, in view of the hyperbolic identity
	
	\begin{equation}\label{eq:hyptrigid}
	\sinh(a)\sinh(b) = \sinh^2\left(\frac{a+b}{2}\right) - \sinh^2\left(\frac{a-b}{2}\right).
	\end{equation}
	
	Our results agree in the suitable limits with those for the modified P\"oschl-Teller potential\cite{Flugge2012,Landau2013}, the Woods-Saxon potential\cite{Landau2013} and the Morse-Feshbach potential\cite{Morse1953}.

\section{Integral identity}
\label{sec:integral}

	The continuum eigenstates states of the Rosen-Morse potential should form an orthogonal basis for the space of scattering states. However, in order to create an integral transform that implements this change of basis we need to evaluate their normalization. Before doing that, we prove a very useful integral identity, which highlights that the generalized Legendre functions have interesting properties of their own, and are not just a simple reparametrization of the hypergeometric function.
	
	We start by defining the function
	
	\begin{equation}\label{eq:cross}
	y(x) = D^{\mu,\eta}_\nu   \frac{dD^{\rho,\gamma}_\sigma}{dx}-D^{\rho,\gamma}_\sigma \frac{dD^{\mu,\eta}_\nu  }{dx}.
	\end{equation}
	
	If we differentiate this expression:
	
	\begin{equation}\label{eq:derivcross}
	\frac{dy}{dx} = D^{\mu,\eta}_\nu   \frac{d^2D^{\rho,\gamma}_\sigma}{dx^2}-D^{\rho,\gamma}_\sigma \frac{d^2D^{\mu,\eta}_\nu  }{dx^2}.
	\end{equation}
	
	Notice we can rearrange \eqref{eq:defeq} as
	
	\begin{equation}\label{eq:rearr}
	\frac{d^2D^{\mu,\eta}_\nu  }{dx^2} = \frac{2x}{1-x^2}\frac{dD^{\mu,\eta}_\nu  }{dx} + \left[\frac{\mu^2+2\mu\eta x + \eta^2}{(1-x^2)^2}-\frac{\nu(\nu+1)}{1-x^2}\right]D^{\mu,\eta}_\nu ,
	\end{equation}
and substituting \eqref{eq:rearr} into \eqref{eq:derivcross} we get
	
	\begin{equation}
	\frac{dy}{dx} - \frac{2x}{1-x^2}y = \left[\frac{\rho^2+\gamma^2-\mu^2-\eta^2+2x(\rho\gamma-\mu\eta)}{(1-x^2)^2}-\frac{\sigma(\sigma+1)-\nu(\nu+1)}{1-x^2}\right]D^{\mu,\eta}_\nu   D^{\rho,\gamma}_\sigma.
	\end{equation}
	
	This is a first order differential equation which can be solved by multiplying by the integrating factor $\exp\left(-\int \frac{2xdx}{1-x^2}\right) = (1-x^2)$. At the end, we obtain
	
	\begin{equation}
	\begin{aligned}
	\frac{d}{dx}&\left[(1-x^2)\left(D^{\mu,\eta}_\nu  \frac{dD^{\rho,\gamma}_\sigma}{dx}-D^{\rho,\gamma}_\sigma \frac{dD^{\mu,\eta}_\nu }{dx}\right)\right] \\
	&=  \left[\frac{\rho^2+\gamma^2-\mu^2-\eta^2+2x(\rho\gamma-\mu\eta)}{1-x^2}-\sigma(\sigma+1)+\nu(\nu+1)\right]D^{\mu,\eta}_\nu  D^{\rho,\gamma}_\sigma.
	\end{aligned}
	\end{equation}
	
	From all of this, we have proved
	
	\begin{equation}\label{eq:identity}
	\begin{aligned}
	\int_a^b&\left[\frac{\rho^2+\gamma^2-\mu^2-\eta^2+2x(\rho\gamma-\mu\eta)}{1-x^2}-\sigma(\sigma+1)+\nu(\nu+1)\right]D^{\mu,\eta}_\nu  D^{\rho,\gamma}_\sigma dx \\
	&= \left[(1-x^2)\left(D^{\mu,\eta}_\nu  \frac{dD^{\rho,\gamma}_\sigma}{dx}-D^{\rho,\gamma}_\sigma \frac{dD^{\mu,\eta}_\nu }{dx}\right)\right]\bigg|_a^b.
	\end{aligned}
	\end{equation}

\section{Normalization and spectral measure}
\label{sec:norm}

Recall the mapping between the wave number $k$ and our generalized Legendre functions
	
	\begin{equation}\label{eq:wavefn}
	\psi_k(x) = \begin{cases}
		D^{-\frac{\sqrt{4\beta-k^2}}{2} + \frac{ik}{2},-\frac{\sqrt{4\beta-k^2}}{2} - \frac{ik}{2}}_\alpha(\tanh x), & |k| \leq 2\sqrt{\beta} \\
		D^{\frac{i\sgn(k)\sqrt{k^2-4\beta}}{2} + \frac{ik}{2},\frac{i\sgn(k)\sqrt{k^2-4\beta}}{2} - \frac{ik}{2}}_\alpha(\tanh x), & |k|>2\sqrt{\beta}
	\end{cases}.
	\end{equation}
	
	Our goal is to evaluate the integral
	
	\begin{equation}
	I = \int_{-\infty}^\infty \psi_p^*(x)\psi_k(x)dx.
	\end{equation}
	
	For this, we use \eqref{eq:identity}. We set $\nu=\sigma=\alpha$ and notice that, for the choice of parameters in \eqref{eq:wavefn}, we have $\mu\eta=\rho\gamma=\beta$, $\mu^2+\eta^2 = 2\beta - k^2$ and $\rho^2+\gamma^2 = 2\beta - p^2$. In this setting, lots of simplifications occur in \eqref{eq:identity} and we obtain
	
	\begin{equation}
	(k^2-p^2)\int_a^b \frac{\bar{D}^{ip}(u)^*\bar{D}^{ik}(u)}{1-u^2}du = \left[(1-u^2)\left(\bar{D}^{ik}\frac{d\bar{D}^{ip*}}{du}-\bar{D}^{ip*} \frac{d\bar{D}^{ik}}{du}\right)\right]\bigg|_a^b,
	\end{equation}
where we have introduced $\bar{D}^{ik}$ as an abreviation for the generalized Legendre functions in \eqref{eq:wavefn}.

	After doing the change of variables $u=\tanh x$:
	
	\begin{equation}
	\begin{aligned}
	(k^2-p^2)\int_a^b &\bar{D}^{ip*}(\tanh x)\bar{D}^{ik}(\tanh x)dx \\
	&= \left(\bar{D}^{ik}(\tanh x)\frac{d}{dx}\bar{D}^{ip*}(\tanh x)-\bar{D}^{ip*}(\tanh x) \frac{d}{dx}\bar{D}^{ik}(\tanh x)\right)\bigg|_a^b.
	\end{aligned}
	\end{equation}
	
	And we may finally express our orthogonality integral as
	
	\begin{equation}\label{eq:intfinal}
	I = \frac{1}{k^2-p^2}\left(\bar{D}^{ik}(\tanh x)\frac{d}{dx}\bar{D}^{ip*}(\tanh x)-\bar{D}^{ip*}(\tanh x) \frac{d}{dx}\bar{D}^{ik}(\tanh x)\right)\bigg|_{-\Lambda}^\Lambda,
	\end{equation}
where we have introduced a cutoff $\Lambda\to\infty$ to regularize the integral.

	We need to subtract two terms which are very similar. Let's focus on the first one. The asymptotics at $\Lambda$ is only nonzero for $|k|>2\sqrt{\beta}$, when it's equal to

	\begin{equation}
	\bar{D}^{ik}(\tanh \Lambda) = 2^{-\eta}e^{i\sqrt{k^2-4\beta}\Lambda},
	\end{equation}
and the derivative is

	\begin{equation}\label{eq:derivinf}
	\frac{d}{dx}\bar{D}^{ip*}(\tanh \Lambda) = -i2^{-\gamma}\sgn(p)\sqrt{p^2-4\beta} e^{-i\sgn(p)\sqrt{p^2-4\beta}\Lambda},
	\end{equation}
so that the total contribution at $\Lambda$ is given by

	\begin{equation}
	T(\Lambda) = -i2^{-\eta-\gamma}\left(\sgn(k)\sqrt{k^2-4\beta}+\sgn(p)\sqrt{p^2-4\beta}\right)e^{i\left(\sgn(k)\sqrt{k^2-4\beta}-\sgn(p)\sqrt{p^2-4\beta}\right)\Lambda}.
	\end{equation}
	
	To make further progress, we need the distributional relations
	
	\begin{equation}
	\lim_{\Lambda \to \infty} \frac{\sin(\Lambda x)}{x} = \pi\delta(x), \qquad \lim_{\Lambda \to \infty} \frac{\cos(\Lambda x)}{x} = 0.
	\end{equation}
	
	They allow us to get rid of the regulator with the substitution

	\begin{equation}\label{eq:deltafn}
	\frac{e^{-i(f(k)-f(p))\Lambda}}{f(k)-f(p)} \to -i\pi\delta(f(p)-f(k)).
	\end{equation}
	
	Using some algebra, when we divide by $k^2-p^2$ we obtain
	
	\begin{equation}
	\frac{T(\Lambda)}{k^2-p^2} = -i2^{-\eta-\gamma}\frac{e^{i\left(\sgn(k)\sqrt{k^2-4\beta}-\sgn(p)\sqrt{p^2-4\beta}\right)\Lambda}}{\sgn(k)\sqrt{k^2-4\beta}-\sgn(p)\sqrt{p^2-4\beta}} \to \pi 2^{-\eta-\gamma}\delta\left(\sgn(k)\sqrt{k^2-4\beta}-\sgn(p)\sqrt{p^2-4\beta}\right).
	\end{equation}
	
	We now use the identity
	
	\begin{equation}
	\delta(f(x)) = \sum_i \frac{\delta(x-x_i)}{|f'(x_i)|},
	\end{equation}
where $x_i$ denotes the zeros of $f(x)$. This is where the $\sgn$ function is seen to be important, because the only zero is when $k=p$. We get, after substituting $\gamma=-\eta$:

	\begin{equation}
	I(\Lambda) = \frac{T(\Lambda)}{k^2-p^2} = \pi \frac{\sqrt{k^2-4\beta}}{|k|}\delta\left(k-p\right).
	\end{equation}
	
	We now look at the more interesting $x=-\Lambda$ asymptotics. It's given by \eqref{eq:limminus}:
	
	\begin{equation}\label{eq:waveminf}
	\bar{D}^{ik}(\tanh -\Lambda) =  2^{-\eta}\left[ \frac{\Gamma(1-\mu-\eta)\Gamma(-\mu+\eta)}{\Gamma(1-\mu+\alpha)\Gamma(-\mu-\alpha)}e^{-(\mu-\eta)\Lambda} + \frac{\Gamma(1-\mu-\eta)\Gamma(\mu-\eta)}{\Gamma(-\alpha-\eta)\Gamma(\alpha+1-\eta)} e^{(\mu-\eta)\Lambda}\right],
	\end{equation}
and the derivative is

	\begin{equation}\label{eq:derivminf}
	\begin{aligned}
\frac{d}{dx}&\bar{D}^{ip*}(\tanh -\Lambda) = 2^{-\gamma}\bigg[ (\rho-\gamma)\frac{\Gamma(1-\rho-\gamma)\Gamma(-\rho+\gamma)}{\Gamma(1-\rho+\alpha)\Gamma(-\rho-\alpha)}e^{-(\rho-\gamma)\Lambda}\\
	&+ (\gamma-\rho)\frac{\Gamma(1-\rho-\gamma)\Gamma(\rho-\gamma)}{\Gamma(-\alpha-\gamma)\Gamma(\alpha+1-\gamma)} e^{(\rho-\gamma)\Lambda}\bigg].
	\end{aligned}
	\end{equation}
	
	When we multiply the terms in \eqref{eq:derivminf}, the crossed terms will have exponential factors like
	
	\begin{equation}
	e^{(\rho-\mu-\gamma+\eta)\Lambda} = e^{-i(p+k)\Lambda}.
	\end{equation}
	
	We get rid of the regulator by doing something analogous to \eqref{eq:deltafn}. We can gobble a $p+k$ term from the denominator of \eqref{eq:intfinal} and write
	\begin{equation}\label{eq:delta}
	\frac{e^{-i(p+k)\Lambda}}{p+k} \to -i\pi\delta(p+k).
	\end{equation}
	
	Notice, though, that once we've gathered all contibutions from the first term of \eqref{eq:intfinal}, the function multiplying it has to be symmetric with respect to the interchange $k\to-p$ and $p\to-k$. However, we would get the same contribution from the second term of \eqref{eq:intfinal}, because it's the same calculation under the same relabeling. Since we need to subtract these two, the factors multiplying $\delta(p+k)$ necessarily vanish, which means we can completely ignore the crossed terms in the multiplication of \eqref{eq:waveminf} and \eqref{eq:derivminf}, and focus only on the uncrossed ones.
	
	The terms are
	
	\begin{equation}
	T_1 = -i2^{-\eta-\gamma}p\frac{\Gamma(1-\mu-\eta)\Gamma(-\mu+\eta)}{\Gamma(1-\mu+\alpha)\Gamma(-\mu-\alpha)}\frac{\Gamma(1-\rho-\gamma)\Gamma(-\rho+\gamma)}{\Gamma(1-\rho+\alpha)\Gamma(-\rho-\alpha)}e^{-i(k-p)\Lambda},
	\end{equation}
	
	\begin{equation}
	T_2 = i2^{-\eta-\gamma}p\frac{\Gamma(1-\mu-\eta)\Gamma(\mu-\eta)}{\Gamma(-\alpha-\eta)\Gamma(\alpha+1-\eta)}\frac{\Gamma(1-\rho-\gamma)\Gamma(\rho-\gamma)}{\Gamma(-\alpha-\gamma)\Gamma(\alpha+1-\gamma)}e^{i(k-p)\Lambda}
	\end{equation}
	
	To compute the second term of \eqref{eq:intfinal} we simply send $k\to-p$ and $p\to-k$. With this transformation, $(k-p)$ stays unchanged, and we have $\mu \to \rho$, $\eta \to \gamma$, and vice-versa. Since the gamma functions are all symmetric with respect to parameter changes, the only thing that changes is the $p$ factor at the beginning, which goes to $-k$. However, since we need to subtract, this gives a $(k+p)$ factor in both $T_1$ and $T_2$, which cancels the one in the denominator in \eqref{eq:intfinal}.
	
	Finally, we gobble the $(k-p)$ factor in the denominator of \eqref{eq:intfinal} to perform a substitution like the one in \eqref{eq:delta}, and at the end we have the total terms at $-\Lambda$:
	
	\begin{equation}
	I_1(-\Lambda) =  -\pi 2^{-\eta-\gamma}\frac{\Gamma(1-\mu-\eta)\Gamma(-\mu+\eta)}{\Gamma(1-\mu+\alpha)\Gamma(-\mu-\alpha)}\frac{\Gamma(1-\rho-\gamma)\Gamma(-\rho+\gamma)}{\Gamma(1-\rho+\alpha)\Gamma(-\rho-\alpha)}\delta(k-p),
	\end{equation}
	
	\begin{equation}
	I_2(-\Lambda) = -\pi 2^{-\eta-\gamma}\frac{\Gamma(1-\mu-\eta)\Gamma(\mu-\eta)}{\Gamma(-\alpha-\eta)\Gamma(\alpha+1-\eta)}\frac{\Gamma(1-\rho-\gamma)\Gamma(\rho-\gamma)}{\Gamma(-\alpha-\gamma)\Gamma(\alpha+1-\gamma)}\delta(k-p),
	\end{equation}
where

	\begin{equation}
	I = I(\Lambda) -I_1(-\Lambda) - I_2(-\Lambda).
	\end{equation}
	
	The parameters $\rho$ and $\gamma$ are obtained by substituting $p$ in either \eqref{eq:scattering} or \eqref{eq:scatpos} and complex conjugating. The delta function forces $p=k$. If we have $k^2<4\beta$, then we should replace $\rho=\eta$ and $\gamma=\mu$. If we have $k^2>4\beta$, we should replace $\rho=-\mu$ and $\gamma=-\eta$. Let's do $k^2>4\beta$ first.
	
	\begin{equation}
	I_1(-\Lambda) =  -\pi \frac{\Gamma(1-\mu-\eta)\Gamma(-\mu+\eta)}{\Gamma(1-\mu+\alpha)\Gamma(-\mu-\alpha)}\frac{\Gamma(1+\mu+\eta)\Gamma(\mu-\eta)}{\Gamma(1+\mu+\alpha)\Gamma(\mu-\alpha)}\delta(k-p),
	\end{equation}
	
	\begin{equation}
	I_2(-\Lambda) = -\pi\frac{\Gamma(1-\mu-\eta)\Gamma(\mu-\eta)}{\Gamma(-\alpha-\eta)\Gamma(\alpha+1-\eta)}\frac{\Gamma(1+\mu+\eta)\Gamma(-\mu+\eta)}{\Gamma(-\alpha+\eta)\Gamma(\alpha+1+\eta)}\delta(k-p).
	\end{equation}
	
	Applying the shift and reflection identities of the gamma function, we simplify:
	
	\begin{equation}
	I_1(-\Lambda) =  -\pi \frac{\mu+\eta}{-\mu+\eta}\frac{\sin(\pi(\mu-\alpha))\sin(\pi(-\mu-\alpha))}{\sin(\pi(\mu-\eta))\sin(\pi(\mu+\eta))}\delta(k-p),
	\end{equation}
	
	\begin{equation}
	I_2(-\Lambda) =  -\pi \frac{\mu+\eta}{-\mu+\eta}\frac{\sin(\pi(\eta-\alpha))\sin(\pi(-\eta-\alpha))}{\sin(\pi(\mu-\eta))\sin(\pi(\mu+\eta))}\delta(k-p).
	\end{equation}
	
	And trigonometric identities:
	
	\begin{equation}
	I_1(-\Lambda) =  -\pi \frac{\sqrt{k^2-4\beta}}{|k|}\frac{\sin^2(\pi \alpha)+\sinh^2\left(\frac{\pi}{2}(\sqrt{k^2-4\beta}+k)\right)}{\sinh(\pi k)\sinh(\pi \sqrt{k^2-4\beta})}\delta(k-p),
	\end{equation}
	
	\begin{equation}
	I_2(-\Lambda) =  -\pi \frac{\sqrt{k^2-4\beta}}{|k|}\frac{\sin^2(\pi \alpha)+\sinh^2\left(\frac{\pi}{2}(\sqrt{k^2-4\beta}-k)\right)}{\sinh(\pi k)\sinh(\pi \sqrt{k^2-4\beta})}\delta(k-p),
	\end{equation}
	
	When evaluating $I(\Lambda)-I_1(-\Lambda)-I_2(-\Lambda)$ for this region, we can simplify the sum using \eqref{eq:hyptrigid}.
	
	Now the region $k^2<4\beta$:
	
	\begin{equation}
	I_1 = I_2 =  -\pi 2^{-\eta-\mu}\frac{\Gamma(1-\mu-\eta)\Gamma(-\mu+\eta)}{\Gamma(1-\mu+\alpha)\Gamma(-\mu-\alpha)}\frac{\Gamma(1-\eta-\mu)\Gamma(-\eta+\mu)}{\Gamma(1-\eta+\alpha)\Gamma(-\eta-\alpha)}\delta(k-p),
	\end{equation}
	
	Substituting $\mu$ and $\eta$ from \eqref{eq:scattering}:
	
	\begin{equation}
	I_1 = -\frac{\pi^2 2^{-\sqrt{4\beta-k^2}}}{k\sinh(\pi k)}\frac{\Gamma\left(1+\sqrt{4\beta-k^2}\right)^2}{\left|\Gamma\left(1+\alpha+\frac{\sqrt{4\beta-k^2}}{2}+\frac{ik}{2}\right)\right|^2\left|\Gamma\left(-\alpha+\frac{\sqrt{4\beta-k^2}}{2}+\frac{ik}{2}\right)\right|^2}\delta(k-p)
	\end{equation}
	
	Therefore, if we let
	
	\begin{equation}\label{eq:measure}
	w_{\alpha,\beta}(k) = \begin{cases}
		2\pi \frac{\sqrt{k^2-4\beta}}{|k|}\frac{\sin^2(\pi \alpha)+\sinh^2\left(\frac{\pi}{2}(\sqrt{k^2-4\beta}+k)\right)}{\sinh(\pi k)\sinh(\pi \sqrt{k^2-4\beta})}, & |k| > 2\sqrt{\beta} \\
		\frac{2\pi^2 2^{-\sqrt{4\beta-k^2}}}{k\sinh(\pi k)}\frac{\Gamma\left(1+\sqrt{4\beta-k^2}\right)^2}{\left|\Gamma\left(1+\alpha+\frac{\sqrt{4\beta-k^2}}{2}+\frac{ik}{2}\right)\right|^2\left|\Gamma\left(-\alpha+\frac{\sqrt{4\beta-k^2}}{2}+\frac{ik}{2}\right)\right|^2}, & |k| \leq 2\sqrt{\beta}
	\end{cases}
	\end{equation}
we can write the normalization of the generalized Legendre functions as

	\begin{equation}\label{eq:ortho}
	\int_{-\infty}^\infty \bar{D}_{\alpha,\beta}^{ip*}(\tanh x)\bar{D}_{\alpha,\beta}^{ik}(\tanh x)dx = w_{\alpha,\beta}(k)\delta(k-p).
	\end{equation}
	
	We can setup something analogous to the Fourier transform. If we suppose an arbitrary scattering state can be expanded into Legendre functions, we write
	
	\begin{equation}\label{eq:fx}
	f(x) = \int_{-\infty}^\infty F(k)\bar{D}^{ik}(\tanh x)dk
	\end{equation}
	
	In view of \eqref{eq:ortho}, we can find $F(k)$ by multiplying by another Legendre function and integrating:
	
	\begin{equation}\label{eq:fk}
	\int_{-\infty}^\infty f(x)\bar{D}_{\alpha,\beta}^{ik*}(\tanh x)dx = F(k)w_{\alpha,\beta}(k)
	\end{equation}
	
	But substituting \eqref{eq:fk} on \eqref{eq:fx} we have
	
	\begin{equation}\label{eq:inttheorem}
	f(x) = \int_{-\infty}^\infty \int_{-\infty}^\infty \frac{\bar{D}_{\alpha,\beta}^{ik*}(\tanh y)\bar{D}_{\alpha,\beta}^{ik}(\tanh x)}{w_{\alpha,\beta}(k)}f(y)dy dk
	\end{equation}
	
	But, since we know that
	
	\begin{equation}
	f(x) = \int_{-\infty}^\infty \delta(x-y)f(y)dy,
	\end{equation}
we must identify the transform on \eqref{eq:inttheorem} as a delta function, and we obtain the resolution of the identity for the scattering states:
	
	\begin{equation}\label{eq:resolution}
	\int_{-\infty}^\infty \frac{\bar{D}_{\alpha,\beta}^{ik*}(\tanh y)\bar{D}_{\alpha,\beta}^{ik}(\tanh x)}{w_{\alpha,\beta}(k)}dk = \delta(x-y)
	\end{equation}
	
	We also note that, when $\beta=0$, \eqref{eq:measure} reduces to 
	
	\begin{equation}
	w_{\alpha, 0}(k) = 2\pi\left(1+\frac{\sin^2(\pi\alpha)}{\sinh^2(\pi k)}\right),
	\end{equation}
and in this limit our functions reduce to regular associated Legendre functions like

	\begin{equation}
	\bar{D}_{\alpha,0}^{ik}(\tanh x) = \Gamma(1-ik)P^{ik}_\alpha(\tanh x),
	\end{equation}
so that when we substitute this on \eqref{eq:resolution}, we have the resolution of the identity for scattering states of the symmetric potential with respect to Legendre functions:

	\begin{equation}
\int_{-\infty}^\infty \frac{kdk\sinh(\pi k)P_\alpha^{-ik}(\tanh y)P_\alpha^{ik}(\tanh x)}{2(\sinh^2(\pi k)+\sin^2(\pi\alpha))} = \delta(x-y),
	\end{equation}
which matches with the one obtained by means of path integrals\cite{Grosche1998}.

\section{Conclusions}
\label{sec:concl}

In this article, we have generalized associated Legendre functions and shown how they naturally describe states of the Rosen-Morse potential, allowing for a completely classical solution of the system, from which we can obtain reasonably simple expressions for the reflection and transmission coefficients and the normalization of the continuum spectrum, all of which reduced in the suitable limit to the ones reported in the literature. The integral identity obtained in this work shows that these functions are interesting in their own right, and more investigation of their properties should give more insight into exactly solvable potentials.

\bibliography{rmscat}

\end{document}